%% file: main.tex
\documentclass[conference]{IEEEtran}
\IEEEoverridecommandlockouts
\usepackage{cite}
\usepackage{amsmath,amssymb,amsfonts}
\usepackage{graphicx}
\usepackage{textcomp}
\usepackage{xcolor}
\def\BibTeX{{\rm B\kern-.05em{\sc i\kern-.025em b}\kern-.08em
    T\kern-.1667em\lower.7ex\hbox{E}\kern-.125emX}}

\usepackage{cite}
\usepackage{tabularx}
\usepackage{hyperref}
\usepackage{cleveref}
\usepackage{xcolor}
\usepackage{ulem}
\usepackage{algorithm}
\usepackage{algpseudocode}
\usepackage{booktabs}
\usepackage{multirow}
\usepackage{graphicx}
\usepackage{subcaption}
\usepackage{enumitem}
\usepackage{tikz}

\newcommand{\del}[1]{\textcolor{red}{}}
\newcommand{\pn}{\textit{Platinum}}

\begin{document}



\title{Platinum: \underline{P}ath-Adaptable \underline{L}UT-Based \underline{A}ccelerator \underline{T}ailored for Low-Bit Weight Matrix Multiplication
}



\author{
Haoxuan~Shan, Cong~Guo,~\IEEEmembership{Member,~IEEE,} Chiyue~Wei, Feng~Cheng, Junyao~Zhang, \\
Hai~(Helen)~Li,~\IEEEmembership{Fellow,~IEEE,} Yiran~Chen,~\IEEEmembership{Fellow,~IEEE} \\
\textit{Department of Electrical and Computer Engineering, Duke University}\\
\textit{\{haoxuan.shan, cong.guo, chiyue.wei, feng.cheng, junyuao.zhang, hai.li, yiran.chen\}@duke.edu}
\thanks{This work was supported in part by NSF grants 2328805 and 2112562. The authors thank the anonymous reviewers for their constructive feedback. Cong Guo is the corresponding author of this paper.}
}


\maketitle

\input{tex/0_abstract}


\input{tex/1_introduction}

\input{tex/2_background}
\input{tex/3_methodology}
\input{tex/4_system_design}
\input{tex/5_evaluation}
\input{tex/6_conclusion}

\newpage

\bibliographystyle{IEEEtran}
\bibliography{refs}


\end{document}

%% file: tex/0_abstract.tex
\begin{abstract}

The rapid scaling of large language models demands more efficient hardware. 
Quantization offers a promising trade-off between efficiency and performance. With ultra-low-bit quantization, there are abundant opportunities for results reuse, and thus it can be boosted with lookup tables (LUTs) based acceleration. 
However, existing LUT-based methods suffer from computation and hardware overheads for LUT construction, and rely solely on bit-serial computation, which is suboptimal for ternary-weight networks.
We propose \pn{}, a lightweight ASIC accelerator for integer weight mixed-precision matrix multiplication (mpGEMM) using LUTs. \pn{} reduces LUT construction overhead via offline-generated construction paths and supports both general bit-serial and optimized ternary-weight execution through adaptive path switching.
On BitNet b1.58-3B, \pn{} achieves up to 73.6×, 4.09×, and 2.15× speedups over SpikingEyeriss, Prosperity, and 16-thread T-MAC (CPU), respectively, along with energy reductions of 32.4×, 3.23×, and 20.9×, all within a 0.96mm$^2$ chip area.
This demonstrates the potential of LUT-based ASICs as efficient, scalable solutions for ultra-low-bit neural networks on edge platforms.



\end{abstract}

%% file: tex/1_introduction.tex
\section{Introduction}

The size of deep neural networks (DNNs), particularly large language models (LLMs), continues to grow rapidly~\cite{hugo2023llama, touvron2023llama2, shoeybi2019megatron}, leading to increased energy consumption and computational latency. Among core operations in LLMs, general matrix multiplication (GEMM) dominates both fully connected and attention layers. As model sizes scale, GEMM’s computational burden grows proportionally, creating major deployment challenges.

Quantization has emerged as a promising solution by reducing computation and memory overhead with minimal accuracy loss. Numerous studies show that aggressive low-bit quantization yields substantial efficiency gains~\cite{cong2022ant, guo2023olive, peng2023fp8lm,cheng2025ecco, guo2025survey, haoxuan2025neuromorphic}. Instead of uniformly quantizing weights and activations, applying low-bit quantization to weights alone has shown promise~\cite{lin2023awq, frantar2022gptq, hongyu2023bitnet, shuming2024b158}, motivating the need for accelerating mixed-precision GEMM (mpGEMM). For example, BitNet-b1.58~\cite{shuming2024b158} uses ternary weights ({-1, 0, 1}) to balance efficiency and accuracy.

Interestingly, ultra-low-bit quantization enables LUT-based acceleration strategies, which store precomputed results for reuse. This approach is well-suited for LLMs, where large hidden dimensions allow reusing between thousands of weight multiplications per input. 
Combined with bit-serial approaches~\cite{jianyu2024tmac, chen2025bitmod,judd2016stripes_bitserial,guo2025transitive}, LUT methods can efficiently support integer-based quantization mpGEMM. As a result, LUT-based acceleration~\cite{yongkweon2020biqgemm, zhiwen2024luttensorcore, gunho2024lutgemm, jianyu2024tmac, jinheng2025bitnetcpp, kim2025slimllama,wei2025phi} has recently gained attention for efficient LLM inference.

\begin{figure}[t]
\centering
\includegraphics[width=0.8\linewidth]{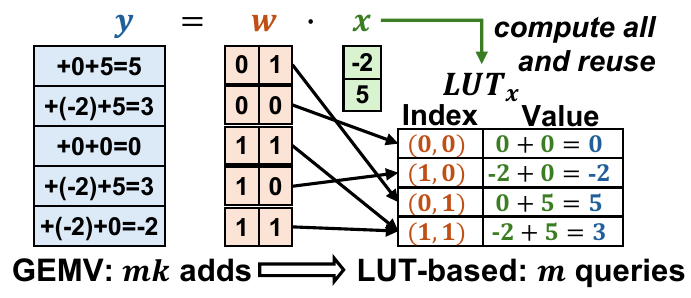}
\vspace{-2mm}
\caption{Comparison between original binary matrix vector multiplication (GEMV) and LUT-based optimization using a binary weight matrix $\mathbf{w}$ of shape $(m, k)$ and input vector $\mathbf{x}$ of shape $(k, n)$, where $m=5, n=1, k=2$. LUT-based optimization reduces computation by a factor of $k$.}
\vspace{-5mm}
\label{fig:binary_lut_mpgemm}
\end{figure}

Precomputing all potential LUTs and loading them at runtime is prohibitively costly. Most existing methods build LUTs at runtime for incoming activations and query with weights. Minimizing LUT construction overhead is thus critical. Prosperity~\cite{wei2025prosperity} addresses this by using “shortcuts” that reuse intermediate entries during construction.
However, its dynamic scheduling incurs high area and energy overhead—unnecessary for models like BitNet with uniformly distributed weights that utilize most LUT entries.

Moreover, existing ASICs typically use bit-serial execution for general weight precision, introducing redundancy for bitwidth like ternary weights. Bit-serial encoding for ternary numbers uses 2 bits, exceeding the optimal 1.58 bits ($\log_23$), and incurs overhead from merging partial sums. In contrast, ternary LUTs directly represent final results, eliminating this cost. Experiments show over 1.3× improvement with ternary LUTs over binary LUTs for ternary weights, highlighting optimization potential for specific bitwidths.

To address these gaps, we propose \textbf{\pn{}}, a \underline{P}ath-adaptable \underline{L}UT-based \underline{A}ccelerator \underline{T}ailored for Low-bit weights mpGEMM. Our key contributions include:
\begin{itemize}[leftmargin=*]
    \item We designed \pn{}, a novel LUT-based accelerator for mpGEMM with integer weights, leveraging a disaggregated path-based LUT construction framework to reduce LUT generation cost and minimize hardware overhead.
    \item The accelerator supports integer-weight mpGEMM via bit-serial execution. By switching the construction path, it can be reconfigured for optimized execution at specific weight precision such as ternary.
    \item The architecture is optimized for parallelism and dataflow efficiency, and is lightweight and modular for edge deployment, with a chip area of just $0.96\,\text{mm}^2$.
    \item \pn{} achieves up to $73.6\times$ speedup and $32.4\times$ energy savings over state-of-the-art baselines on BitNet b1.58 3B, one of the most representative ternary weight models.
\end{itemize}



\begin{figure}[b]
\centering
\includegraphics[width=1.0\linewidth]{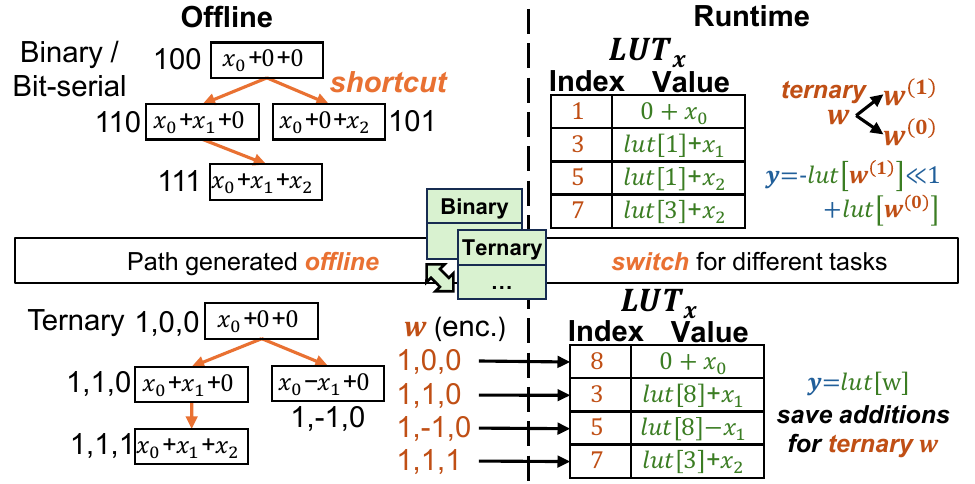}
\caption{\pn{} leveraging programmable construction path to support both bit-serial LUT-based and ternary LUT-based mpGEMM. The path generation is disaggregated to offline to reduce runtime overhead. Refer to \Cref{sec:ternary_lut} for more details of ternary LUT benefits for ternary weights. }
\label{fig:teaser2}
\end{figure}


%% file: tex/2_background.tex

\section{Background}

\textbf{Bit-serial LUT-based mpGEMM.}
Bit-serial execution decomposes a $b$-bit integer weight into $b$ binary matrices $w^{(0)},...,w^{(b-1)}$, and computes $y = wx$ by summing $2^iw^{(i)}x$. For a given input $x$, all $w^{(i)}$ share the same LUT, increasing reuse. In \Cref{fig:binary_lut_mpgemm}, each output element in the product between a binary matrix and input vector has at most $2^k$ unique values, where $k$ is the input vector length. These values can be precomputed and then queried, replacing $mk$ additions with $m$ LUT accesses and $(k-1)2^k$ additions for LUT construction.
When $m$ is large and LUT access cost is lower than $k$ additions, this method reduces runtime computation. LLMs naturally satisfy the first condition due to large hidden dimensions, and the second is met by choosing a large $k$ while ensuring LUTs still reside in cache or on-chip buffers.

\textbf{Tradeoffs in LUT construction.}
Precomputing all LUTs offline is impractical. For example, buffering all LUTs for 8-bit activations with $k=2$ already require 4GB of memory, which is a high cost. As a result, most works build LUTs at runtime for each input vector. However, constructing binary LUTs requires $(k-1)2^k$ additions if each entry is computed independently, which grows exponentially.
To mitigate the cost, prior works~\cite{wei2025prosperity, yongkweon2020biqgemm} introduce shortcut-based construction. Prosperity~\cite{wei2025prosperity} dynamically detects shortcuts between LUT entries to reduce cost, which is effective when only a few entries are accessed. However, dynamic construction incurs significant overhead—its runtime shortcut scheduling modules account for 24\% of chip area and 32.3\% of total power. 
For models like BitNet-b1.58, whose uniformly distributed ternary weights lead to high LUT utilization, runtime path scheduling is thus unnecessary. For instance, only 1.16\% of LUT entries remain unused when tiling the $M$-dimension with size 1080.

In contrast, our accelerator, \pn{}, eliminates runtime overhead by disaggregating LUT construction into offline path generation and lightweight online path-based construction (\Cref{fig:teaser2}). This removes the need for dynamic scheduling hardware, reducing area and energy costs while enabling more processing elements (PEs) for higher throughput.
Moreover, the reprogrammable path design allows \pn{} to support both bit-serial and ternary LUTs. This enables optimized execution for ternary-weight networks, while maintaining compatibility with general weight precisions.

%% file: tex/3_methodology.tex
\section{Methodologies}


\subsection{\pn{} Overview}\label{sec:architecture}

\pn{} (\Cref{fig:b158lut_arch}) comprises $L$ Platinum Processing Elements (PPEs), aggregators, high-bandwidth on-chip buffers, and Special Functional Units (SFUs). 
In each single computation round, it first constructs $L$ LUTs based on input with shape $(Lc,1)$ and then sequentially queries $(m,Lc)$ weights and aggregates the results into the output buffer. The complete algorithm is shown in \Cref{alg:single_computation_round}. $c$ represents the chunk size (vector length $k$) for constructing LUT. The chunk size is set to 5 for ternary weights to align with the weight encoding (\Cref{sec:ternary_lut}), which requires LUTs with 128 entries. 

PPEs are responsible for LUT construction and query. Each includes a controller, adders, and a dedicated LUT buffer. 
Aggregators share the adders in PPEs and work alongside additional adders to form a pipelined adder tree for efficient aggregation and accumulation during the Reduction stage. The rationale for including these extra adders is discussed in \Cref{sec:utilization_improvements}.
LUT buffers are implemented using on-chip SRAM.
Each entry is 8-bit to align with Bitnet's 8-bit activation. Given that the LUT occupies a relatively small portion of the overall chip area, using higher precision (i.e., more bits per entry) is also feasible if needed.
Each LUT buffer includes a read-write port and an additional read-only port to prevent pipeline stalls. Both ports are used simultaneously during the query phase to maximize throughput.
The remaining buffers include storage for weights, inputs, outputs, and build paths. They are divided into banks to maximize throughput. These buffers are accessed sequentially during execution, benefiting from prefetching. 
SFUs are included to support operations beyond mpGEMM, such as vector multiplications and activation functions. Since this work focuses on GEMM, it only serves as a hardware overhead for fair comparison with baselines.


\begin{algorithm}[t]
    \caption{Single Computation Round}
    \label{alg:single_computation_round}
    \begin{algorithmic}[1]
        \State \textbf{Input:} Weights $(m, Lc)$, Inputs $(Lc, 1)$, Outputs $(m, 1)$
        \State \textbf{Output:} Outputs $(m, 1)$
        \State \textbf{Hardware:} Platinum Processing Elements (PPE), Aggregator (AGG)
        
        \State \textbf{function} FLIP($v$,$s$)
        \textbf{return} $\text{if } s \text{ then } -v \text{ else } v$
        
        \Function{ppe.Query}{$index$}
            \State \textbf{return} \Call{Flip}{$LUT[index[6:0]]$, $index[7]$}
        \EndFunction
        
        \Function{agg.Reduce}{$vals$, $output$}
            \State $sum \gets$ \Call{agg.aggregate}{vals}
            \State $output \gets output + sum$
        \EndFunction
        
        \Function{Round}{$inputs$, $weights$, $outputs$}
            \State \Call{ppe.Construct}{$inputs$} (\Cref{alg:lut_construction})
            \For{$i = 0$ to $m-1$}
                \State $vals \gets$ \Call{ppe.query}{$weights[i]$}
                \State \Call{agg.reduce}{$vals$, $outputs[i]$}
            \EndFor
        \EndFunction
        
    \end{algorithmic}
\end{algorithm}

\input{figtex/3_arch}

\begin{figure}[t]
\centering
\includegraphics[width=\linewidth]{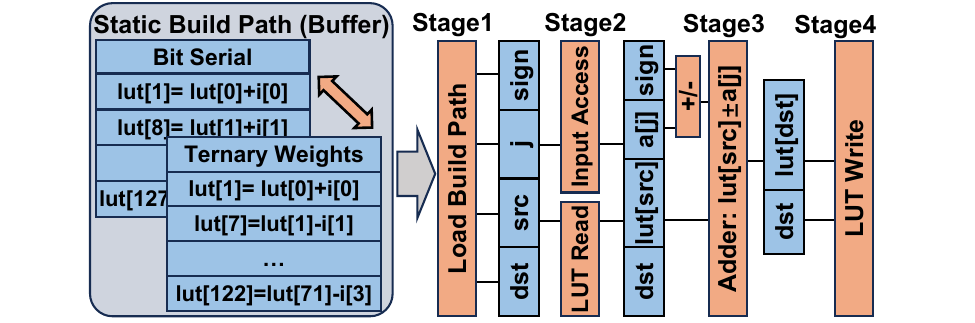}
\vspace{-6mm}
\caption{Four-stage construction pipeline with build path. }
\label{fig:build_pipeline}
\vspace{-6mm}
\end{figure}

\pn{} employs a disaggregated path-based LUT construction scheme (\Cref{alg:lut_construction}). The build path, generated offline, represents shortcut paths as illustrated in \Cref{fig:teaser2} and is sequentially loaded during the LUT construct stage of runtime until a "Finish" token is reached. Each shortcut path triggers the computation of a new LUT entry by combining an existing entry with an input element.
A dedicated four-stage construction pipeline (\Cref{fig:build_pipeline}) is implemented to support runtime construction and minimize construction stalls. 
At each cycle, the pipeline fetches a path entry from the build path buffer and sends to PPEs. The controller in PPEs extracts the source and destination addresses, input index, and sign bit. In the second stage, the required values are read from the LUT and input buffers. These values are combined—either added or subtracted—in the third stage, and the result is written back in the final stage.


\begin{algorithm}[htbp]
    \caption{Path-Based LUT Construction}
    \label{alg:lut_construction}
    \begin{algorithmic}[1]
        \State \textbf{Parameters:} Chunk size $c$, Table size $S$
        \State \textbf{Input:} Input activations $a_1, \dots, a_c$; build\_path[$S$] 
        \State \textbf{Output:} Look-up Table LUT[$S$]
        \Function{LUT\_Construction}{$a_1, \dots, a_c$}
            \State Initialize LUT[0] $\gets 0$, PC $\gets 0$
            \While{build\_path[PC] is not ``Finish''}
                \State path $\gets$ build\_path[PC]
                \State dst, src, $j$, sign $\gets$ path
                \State LUT[dst] $\gets$ LUT[src] $+ \text{Flip}(a_j, sign)$
                \State PC $\gets$ PC $+ 1$
            \EndWhile
            \State \Return LUT
        \EndFunction
    \end{algorithmic}
\end{algorithm}



\subsection{Offline Path Generation with MST}\label{sec:b158lut_build_path}

Previous designs construct each LUT entry in parallel by directly computing from the inputs. 
While this strategy is well-suited for GPUs, which benefit from abundant parallelism and rich computational resources, it introduces significant overhead in compact accelerators with limited on-chip resources. 
This overhead is tolerable for small chunk sizes but a more efficient construction method is desirable for larger chunk sizes and ternary weights. 
BIQGEMM \cite{yongkweon2020biqgemm} proposed a dynamic programming approach for binary weights, obtaining each LUT entry with just a left shift and one addition. 
Prosperity \cite{wei2025prosperity} leveraged "shortcut" to maximize the reuse of entries already computed. 
Inspired by these ideas, we propose a new method from a graph-theoretic perspective with minimal construction cost. 

The LUT construction can be formalized as a directed hypergraph. 
Each node represents a LUT entry, and each hyperedge represents a computation. For example, a hyperedge $e = (\{n_1, n_2\}, n_3, \texttt{add\_cost})$ denotes an addition operation $n_3 = n_1 + n_2$. Given a set of source nodes $\{s_j\}$, a valid build path is a subset of hyperedges that connects all other nodes to the source set. The optimal build path minimizes the total cost and corresponds to a minimum spanning tree (MST) in the hypergraph.
In our setting, the source node is $lut_0$, representing zero, and the operations are restricted to additions/subtractions of input entry, which are reversible. This allows us to convert the hypergraph into a standard undirected graph and apply classical MST algorithms such as Prim’s algorithm \cite{prim1957shortest} for efficient construction. 
The resulting MST yields a build path with operations of the form $(\texttt{dst}, \texttt{src}, j, \texttt{flip})$, representing $lut[\texttt{dst}] = lut[\texttt{src}] \pm a_j$. The build order is naturally topologically sorted from the MST, ensuring correct data dependencies. Notably, for $c = 5$, the shortest Read-After-Write (RAW) dependency distance exceeds the number of pipeline stages, eliminating the need for additional hardware hazard handling.
Combined with LUT size reduction via symmetry, our MST-based approach reduces the number of additions by $\sim10\times$ at $c = 5$, compared to naive LUT construction. 


The generated build path is stored in on-chip buffer and retrieved by the four-stage construction pipeline at runtime. By generating separate build paths for bit-serial and ternary LUT, and dynamically switching paths during execution (\Cref{fig:build_pipeline}), our accelerator efficiently supports multiple weight precisions, including GEMM in attention layers.

\input{tex/3_2_methodology_ternary}

%% file: figtex/3_arch.tex
\begin{figure}[t]
\centering
\vspace{-3mm}
\includegraphics[width=\linewidth]{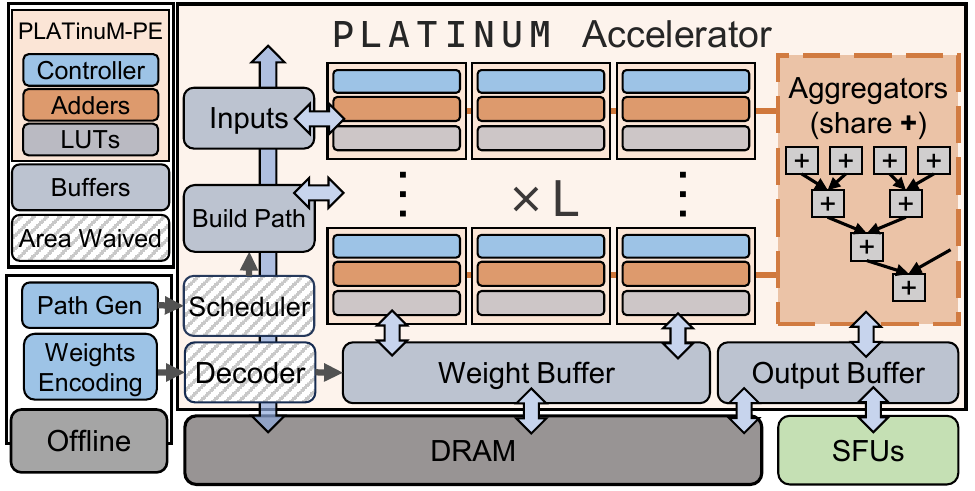}
\caption{Architecture of \pn{} Processor}
\label{fig:b158lut_arch}
\vspace{-5mm}
\end{figure}

%% file: tex/3_2_methodology_ternary.tex
\subsection{LUT Method Tailored for Ternary Weights mpGEMM}\label{sec:ternary_lut}

In this section, we present a ternary LUT-based mpGEMM approach and demonstrate why it outperforms bit-serial LUT methods when handling ternary weights.
For ternary mpGEMM with weights of size $M \times K$ and input of size $K \times N$, naive computation requires $MKN$ additions. Subtractions are counted as additions due to negligible sign-flip cost. The bit-serial LUT method reduces this cost by approximately $c/2$ when $M$ is large, as shown in \Cref{equ:number_of_addition_with_binary_lut}. 
For each $c$-element input chunk, $c2^c$ additions are required to construct the LUT, followed by $2M$ LUT queries and $M$ additions to merge results. This process repeats $\lceil K/c \rceil N$ times. An additional $M(\lceil K/c \rceil - 1)N$ additions are needed to accumulate partial results. 

\vspace{-1em}
\begin{equation}
\small
    \label{equ:number_of_addition_with_binary_lut}
    \#\text{add}_\text{bs} = \left[\left\lceil \frac{K}{c} \right\rceil c 2^c + M \left\lceil \frac{K}{c} \right\rceil + M \left(\left\lceil \frac{K}{c} \right\rceil - 1\right)\right] N
\end{equation}

In contrast, the ternary LUT-based method avoids the $M \lceil K/c \rceil N$ term and has lower total cost with proper choice of $c$, as shown in \Cref{equ:number_of_addition_with_lut_based}. Here, LUT construction requires $c3^c$ additions per chunk, and result accumulation is the same as in the bit-serial case. 

\vspace{-2mm}
\begin{equation}
\small
\label{equ:number_of_addition_with_lut_based}
\#\text{add}_{\text{ter}} = \left[\left\lceil\frac{K}{c}\right\rceil c 3^{c} + M \left(\left\lceil\frac{K}{c}\right\rceil - 1\right)\right] N
\end{equation}

We further reduce cost by exploiting symmetry (mirror consolidation) in ternary weights, a technique used in prior works~\cite{zhiwen2024luttensorcore, jianyu2024tmac, jinheng2025bitnetcpp}.
For example, weight vectors $[-1, 1]$ and $[1, -1]$ are symmetric; when multiplied with the same input vector, the output of one can be obtained by negating the other. We store only LUT entries where the leftmost non-zero value is $+1$ (e.g., [0, 1, -1]) and infer the rest by negation, reducing LUT size to $\lceil 3^c / 2 \rceil$.
Combined with our offline path-based LUT construction (\Cref{sec:b158lut_build_path}), this approach reduces LUT construction cost from $c3^c$ to $\lceil 3^c / 2 \rceil$, yielding a $\sim2c\times$ reduction compared to naive ternary LUT construction (\Cref{equ:number_of_addition_with_lut_based_opt}). As shown in \Cref{fig:addition_reduction_over_chunk_size}, our method achieves the lowest addition count across varying chunk sizes for BitNet b1.58 3B.


\begin{equation}
\small
    \label{equ:number_of_addition_with_lut_based_opt}
    \#\text{add}_{\text{platinum}} = \left[\left\lceil\frac{K}{c}\right\rceil \left\lceil\frac{3^c}{2}\right\rceil + M \left(\left\lceil\frac{K}{c}\right\rceil - 1\right)\right] N
\end{equation}

\input{figtex/3_chunk_size}

\begin{figure}[t]
\centering
\includegraphics[width=\linewidth]{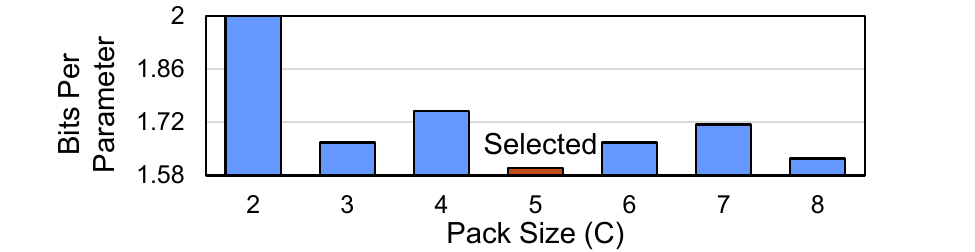}
\vspace{-6mm}
\caption{Average bits per weight w.r.t the pack size $c$.}
\label{fig:weight_encoding_bits}
\vspace{-6mm}
\end{figure}

Compact weight encoding is also critical. Storing each ternary weight as a byte is highly redundant. A common solution, e.g., T-MAC~\cite{jianyu2024tmac}, uses 2-bit encoding, which still far exceeds the optimal 1.58 bits. 
Our work adopts a strategy similar to that used in Bitnet.cpp \cite{jinheng2025bitnetcpp}. Every $c$ ternary weights are packed into a base-3 integer, requiring only $\lceil \log_2 3^c \rceil$ bits. The number is further split into a sign bit and $\lceil \log_2 3^c \rceil - 1$ index bits to preserve symmetry without decoding.
As shown in \Cref{fig:weight_encoding_bits}, encoding overhead is minimized at $c = 5$, achieving 1.6 bits per weight. This fits neatly into a byte and aligns well with typical memory systems. Additionally, we reorder indices based on the construction path to ensure LUT entries are accessed sequentially and no hazard detection is needed for the construction pipeline, forming the final encoded weight stream.
Since the encoding is performed offline, it reduces hardware complexity while avoiding runtime overhead.

%% file: figtex/3_chunk_size.tex

\begin{figure}[t]
\centering
\includegraphics[width=0.95\linewidth]{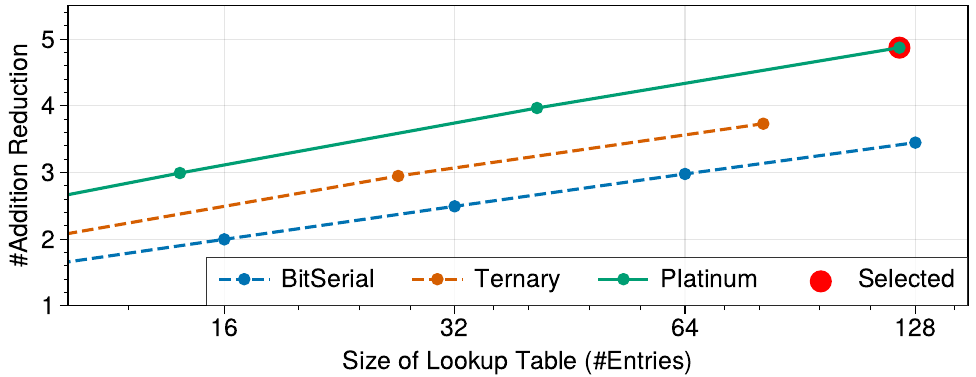}
\vspace{-2mm}
\caption{\#Addition reduction for ternary weights mpGEMM over LUT sizes. Assume $M=1080$. }
\label{fig:addition_reduction_over_chunk_size}
\vspace{-5mm}
\end{figure}

%% file: tex/4_system_design.tex
\section{System-Level Design}\label{sec:system_level_design}


\subsection{Parallelism and Scaling}

There are two primary strategies for improving accelerator throughput: (1) enhancing the processing capacity of each PE, and (2) increasing the number of PEs. 
Our LUT-based GEMM approach exploits the fact that each input vector is reused across many weight vectors, and vice versa-each weight vector serves as a query across multiple input chunks, as also observed in \cite{gunho2024lutgemm}. 
Rather than computing a standard $M \times c$ weight matrix with a $c \times 1$ input, we extend processing to multiple input columns simultaneously, exploiting parallemlism in $N$-dimension. Given $c \times n_{\text{cols}}$ input blocks, we construct a LUT with block size equal to $n_{\text{cols}}$, allowing each query to return a block of $n_{\text{cols}}$ partial sums.
This technique reduces query overhead and overall area consumption compared to a single-column LUT design. Theoretically, increasing $n_{\text{cols}}$ boosts throughput. However, Cacti 7.0 \cite{rajeev2017cacti7} analysis shows diminishing area efficiency beyond $n_{\text{cols}} > 8$. Additionally, for small $N$, large $n_{\text{cols}}$ values cause resource under-utilization. Based on this, we set $n_{\text{cols}} = 8$ in our final design. 

The second scaling axis involves increasing the number of PEs to enable parallel processing along the $K$ and $N$ dimensions. With $L$ PEs, we process $L \cdot c \times n_{\text{cols}}$ inputs in parallel, and stream the resulting partial sums directly to the accumulators, reducing output buffer pressure. However, the choice of $L$ is constrained by memory bandwidth and tiling granularity. Excessive $L$ can result in resource underutilization. We empirically set $L = 52$ to balance throughput and resource efficiency, also facilitating tiling for BitNet-b1.58 models.

\subsection{Utilization Improvements}\label{sec:utilization_improvements}

We observed an imbalance between computation and memory access requirements during the LUT construction, query, and reduction stages. In the construction stage, a stall-free design requires roughly one addition, one LUT read, and one LUT write per cycle, implying one adder per two LUT ports. In contrast, during querying, both LUT ports are used for queries, yielding two partial sums per cycle. This necessitates two adders for reduction to maximize the throughput. This discrepancy results in idle resources: either adders or LUT ports. 
Since the query and reduction stages dominate execution time, it is more beneficial to maximize resource utilization during these stages. To this end, we provision extra adders to fully support the reduction stage's demands. 
It ensures theoretically near 100\% utilization of both LUT ports and an average adder utilization of 90.5\%, resulting in high overall efficiency.

\subsection{Tiling and Stationarity}
Tiling is critical for GEMM acceleration due to the high energy and latency cost of DRAM accesses. 
Large buffers are allocated for weights and outputs to maintain them on-chip while provisioning only minimal input buffering, since it is only accessed during LUT construction.
To determine optimal tiling configurations, we conducted design space exploration on the prefill stages of three BitNet-b1.58 models. Alongside tile sizes, we evaluated various data stationary strategies. The evaluations over configurations across three models are shown in \Cref{fig:dse_tilling}. To strike a balance between latency, energy, and area, we picked m\_tiled=1080, k\_tiled=520, n\_tiled=32 with mnk-stationary, which is marked with red in the figure. We integrate 272KB on-chip SRAM for buffers, together with 52KB LUT, resulting only a total of 324KB. 


\input{figtex/4_tile_size}

%% file: figtex/4_tile_size.tex
\begin{figure}[htbp]
\vspace{-4mm}
\centering
\includegraphics[width=0.9\linewidth]{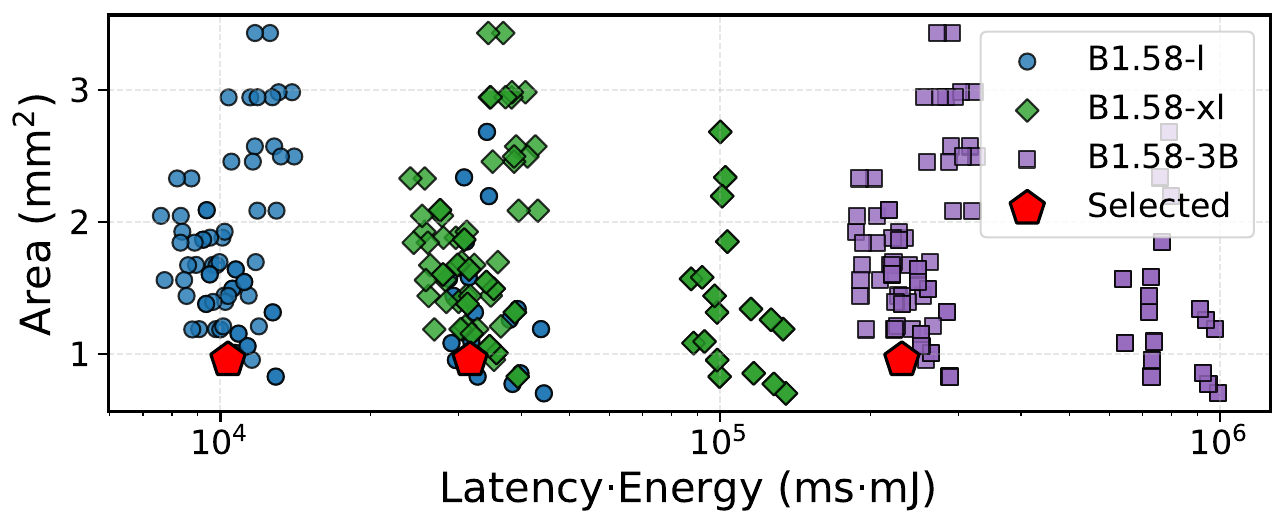}
\vspace{-3mm}
\caption{Design Space Exploration for Tilling Sizes}
\vspace{-4mm}
\label{fig:dse_tilling}
\end{figure}

%% file: tex/5_evaluation.tex
\section{Evaluation}

\input{figtex/5_time}

\input{figtex/5_energy}

\subsection{Experimental Setup}

We evaluate our accelerator using a comprehensive methodology that combines RTL synthesis, memory modeling, DRAM power estimation, and cycle-accurate simulation. 
Both bit-serial based execution (\pn{}-bs) and execution optimized for ternary weights are evaluated (\pn{}). \pn{}-bs adapts binary LUT and $c=7$ to align with the LUT size. 
For benchmarking, we compare our design against SpikingEyeriss~\cite{chen2016eyeriss}, Prosperity~\cite{wei2025prosperity}, and T-MAC~\cite{jianyu2024tmac} under equivalent workloads.
Since SpikingEyeriss~\cite{chen2016eyeriss,surya2020spinalflow, wei2025prosperity}, and Prosperity~\cite{wei2025prosperity} are designed for spiking neural networks, they are evaluated with a bit-serial execution. 
Specifically, for a ternary-weight matrix $W$, each ternary-weight mpGEMM is computed with two passes, handling '1' and '-1' in weights separately. The final result is then obtained by subtracting the results from two passes. 
T-MAC~\cite{jianyu2024tmac} provides a CPU-based implementation. We evaluate it on an Apple M2 Pro laptop with 16 threads, which serves as a strong baseline. 

\input{tabtex/specs}

\noindent \textbf{Model and Kernel Extraction.}
We extract evaluation kernels from the BitNet-b1.58 model suite~\cite{shuming2024b158}, which includes b1.58-l, b1.58-xl, and b1.58-3B models with 700M, 1.3B, and 3B parameters, respectively. These models utilize BitLinear layers as their primary compute blocks. We extract the input ($K$) and output ($M$) feature dimensions from these layers, and vary the product of batch size and sequence length ($N$) to evaluate both the prefill and decode stages of LLM inference. Specifically, we test with $N = 1024$ for the prefill stage and $N = 8$ for the decode stage.

\noindent \textbf{Hardware Modeling.}
The RTL of our \pn{} processing elements is implemented in SystemVerilog. We use Synopsys Design Compiler with ARM’s standard cell library targeting a commercial 28nm technology node and 500 MHz frequency to obtain post-synthesis estimates for area and dynamic/static power.
On-chip SRAM buffer characteristics are modeled using CACTI 7.0~\cite{rajeev2017cacti7}, configured for the same process. 
Off-chip DRAM is modeled using DRAMsim3~\cite{li2020dramsim3} to estimate memory system energy consumption. We followed the settings in \cite{wei2025prosperity} for fair comparison, which employs 64GB DDR4 2133R with 64GB/s as maximum bandwidth. 
We extend the open-source Prosperity~\cite{wei2025prosperity} simulator to support BitNet-b1.58 kernels and to develop a cycle-accurate simulator that captures computation cycles, memory accesses, and PE activities. Using trace outputs and synthesized area/power data, we compute total cycles and energy per kernel. 

\subsection{Area and Power Breakdown}

The chip has an overall size of 0.96 mm$^2$. 
On-chip buffers dominate the overall chip area. Specifically, the buffers for storing weights and activations occupy approximately 65\% of the total area. When including the memory allocated for LUTs, this figure rises to 83.3\%, underscoring the memory-intensive nature of a LUT-based accelerator. The aggregator and PPEs, which carry out the core computations, account for just 15\% of the area. 
The compact footprint of the PPE highlights the area efficiency of compute resources with LUT-based method. 
This benefit becomes even more pronounced when operator area costs increase, such as FP16 adders used for FP16 activations.

When running the prefill workloads of the b1.58-3B model, our accelerator consumes a power of 3.2W. 
DRAM access and weight buffer access are the most power-intensive operations, contributing 53.5\% and 31.6\% of total power, respectively. This highlights the importance of minimizing off-chip memory access and accesses to weight buffers, showing that using compact weight representations is critical to improve energy efficiency. 
Additionally, the LUT buffer exhibits lower power usage compared to the weight buffer and DRAM, indicating that LUT introduces low power overhead in LUT-based computation paradigm. 


\begin{figure}[tp]
\centering
\includegraphics[width=\linewidth]{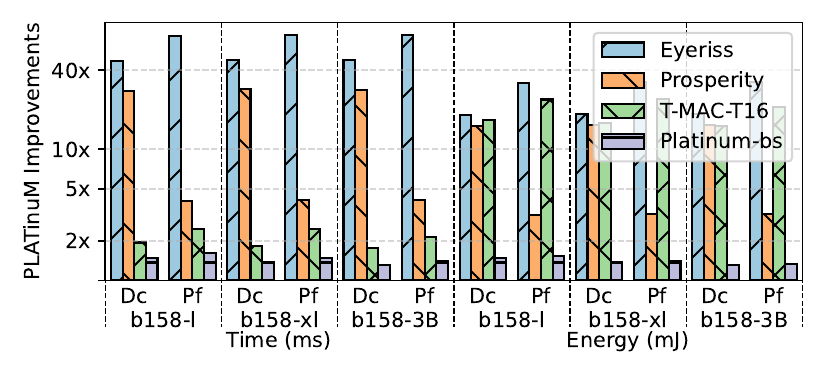}
\vspace{-8mm}
\caption{\pn{} improvements of performance and energy efficiency in Model-level.}
\vspace{-7mm}
\label{fig:model_impr}
\end{figure}

\subsection{Kernel-Level and model-level evaluations}

We evaluate \pn{} and baseline accelerators using representative kernel workloads extracted from BitNet-b1.58, covering both the prefill and decode stages. The results (\Cref{fig:kernel_comparison_latency}, \Cref{fig:kernel_comparison_energy}) demonstrate that \pn{} outperforms state-of-the-art ASIC accelerators and a CPU-based design in terms of both raw performance and energy efficiency. 

Using kernels from the b1.58-3B model configuration (\Cref{fig:model_impr}), \pn{} achieves average speedups of 73.6×, 4.09×, and 2.15× for the prefill stage, and 47.6×, 28.4×, and 1.75× for the decode stage, over SpikingEyeriss, Prosperity, and 16-thread T-MAC, respectively. 
Three key factors contribute to this high performance over the baselines. 
First, by disaggregating "shortcut" path scheduling to offline, it not only benefits from low LUT construction cost, but also minimizes the hardware overhead of runtime scheduling, and thus more area is available for additional PEs.
Second, \pn{} achieves high PE utilization by duplicating adders to fully exploit LUT ports and by choosing $ncols = 8$ per PPE to guarantee utilization under low-$N$ workloads.
In contrast, baseline accelerators like Prosperity suffer from significant underutilization of PEs for decode workloads.
Finally, \pn{} incorporates optimizations specifically for ternary weights, delivering 1.3× and 1.4× speedups over its own bit-serial mode in decode and prefill stages, respectively.

\pn{} also delivers strong energy efficiency, consuming 18.4$\times$, 15.3$\times$, 15.0$\times$, and 1.31$\times$ less energy in the decode stage, and 32.4$\times$, 3.23$\times$, 20.9$\times$, and 1.34$\times$ less in the prefill stage, compared to SpikingEyeriss, Prosperity, T-MAC, and \pn{}-bs, respectively. 
While Prosperity demonstrates high energy efficiency in prefill, \pn{} outperforms it by eliminating runtime scheduler overhead and using compact weight formats to reduce memory access and footprint. Additionally, high parallelism along the $K$ dimension lowers DRAM access frequency of output data, further boosting energy efficiency.


These findings highlight that bit-serial execution on our accelerator already achieves strong performance and efficiency across general weight precisions. Furthermore, optimizations tailored for ternary weights provide an additional performance boost, underscoring the effectiveness of our accelerator for specific weight precision.

%% file: figtex/5_time.tex
\begin{figure*}[htbp]
\centering
\includegraphics[width=1.0\linewidth]{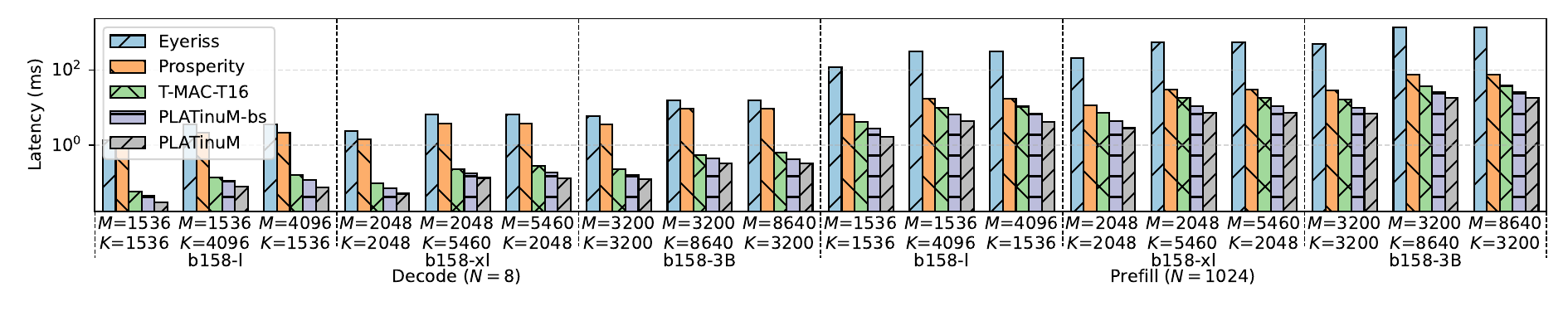}
\vspace{-8mm}
\caption{Comparison of kernel latency across \pn{}, CPU (T-MAC), SpikingEyeriss, and Prosperity.}
\label{fig:kernel_comparison_latency}
\vspace{-5mm}
\end{figure*}

%% file: figtex/5_energy.tex
\begin{figure*}[htbp]
\centering
\includegraphics[width=1.0\linewidth]{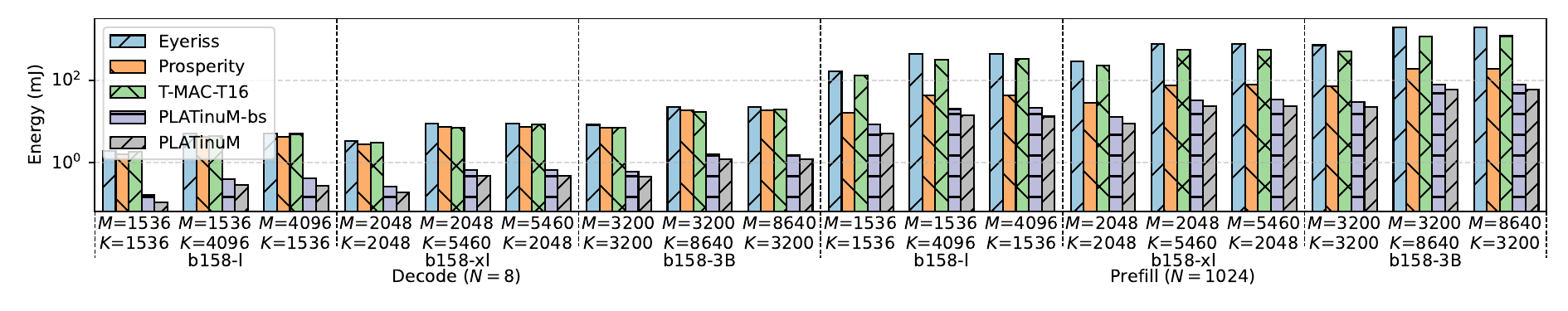}
\vspace{-8mm}
\caption{Comparison of kernel energy across \pn{}, CPU (T-MAC), SpikingEyeriss, and Prosperity.}
\label{fig:kernel_comparison_energy}
\vspace{-6mm}
\end{figure*}

%% file: tabtex/specs.tex
\begin{table}[b]
\centering
\vspace{-4mm}
\caption{Accelerator specifications.}
\vspace{-2mm}
\small
\label{tab:accelerator_specs}
\begin{tabularx}{\linewidth}{l*{4}{>{\centering\arraybackslash}X}}
\toprule
 & Eyeriss\cite{chen2016eyeriss}  & Prosperity\cite{wei2025prosperity} & T-MAC\cite{jianyu2024tmac} & \textbf{\pn \newline Ours} \\
\midrule
Type    &   ASIC    &   ASIC    &   CPU$^\dagger$ &   ASIC    \\
Freq. (MHz)        & 500   & 500   & 3490   & 500    \\
Tech. (nm)       & 28    & 28    & 5     & 28     \\
\# of PEs                    & 168   & 256   & --- & 416   \\
Area (mm$^2$)          & 1.07 & 1.06$^*$ & 289    & 0.955   \\
Throughput$^\ddagger$ (GOP/s)   & 20.8 & 375 &  715  &  1534  \\
\bottomrule
\end{tabularx}
\raggedright{
$^*$ Prosperity is scaled for fair comparison. \\
$^\dagger$ T-MAC: Benchmarked on Apple M2 Pro Laptop.} \\
$^\ddagger$ The number of operations is based on the additions/subtractions for naively compute the b1.58-3B model with $N$=1024. 
\end{table}

%% file: tex/6_conclusion.tex
\section{Conclusion}

In this work, we present \pn{}, a lightweight and energy-efficient ASIC accelerator optimized for low-bit weight matrix multiplication in neural networks. \pn{} employs LUT-based computation with adaptable, offline-generated construction paths to achieve high performance and energy efficiency within a compact hardware footprint. By switching paths and compressing weights offline, it achieves even better efficiency for specific weight precision, such as the ternary weights in BitNet-b1.58 models. Given its efficiency and versatility, \pn{} holds promise for broader adoption in deploying LLM workloads on edge devices.